\documentclass[aps,prl,twocolumn,amsmath,amssymb,superscriptaddress]{revtex4-1}

\usepackage{blindtext}
\usepackage{graphicx}
\usepackage{amssymb}
\usepackage{color}
\usepackage{psfrag}
\usepackage{ifsym}
\usepackage{epstopdf}
\usepackage{soul}
\usepackage{tabularx}

\usepackage{epsfig,color}
\usepackage{amsmath,amssymb}

\usepackage{hyperref}

\newcommand{\bra}[1] {\langle #1 |}
\newcommand{\ket}[1] {| #1 \rangle}
\newcommand{\braket}[2] {\langle #1 | #2 \rangle}
\newcommand{\ketbra}[2] {| #1\rangle \langle  #2 |}

\begin{document}

\title{Measuring Entanglement in a Photonic Embedding Quantum Simulator}

\author{J. C. Loredo}\email[Corresponding author:~]{juan.loredo1@gmail.com}
\affiliation{Centre for Engineered Quantum Systems, Centre for Quantum Computer and Communication Technology, School of Mathematics and Physics, University of Queensland, Brisbane, Queensland 4072, Australia}
\author{M. P. Almeida}
\affiliation{Centre for Engineered Quantum Systems, Centre for Quantum Computer and Communication Technology, School of Mathematics and Physics, University of Queensland, Brisbane, Queensland 4072, Australia}
\author{R. Di Candia}
\affiliation{Department of Physical Chemistry, University of the Basque Country UPV/EHU, Apartado 644, 48080 Bilbao, Spain}
\author{J. S. Pedernales}
\affiliation{Department of Physical Chemistry, University of the Basque Country UPV/EHU, Apartado 644, 48080 Bilbao, Spain}
\author{J. Casanova}
\affiliation{Institut f\"ur Theoretische Physik, Albert-Einstein-Allee 11, Universit\"at Ulm, D-89069 Ulm, Germany }
\author{E. Solano}
\affiliation{Department of Physical Chemistry, University of the Basque Country UPV/EHU, Apartado 644, 48080 Bilbao, Spain}
\affiliation{IKERBASQUE, Basque Foundation for Science, Maria Diaz de Haro 3, 48013 Bilbao, Spain}
\author{A. G. White}
\affiliation{Centre for Engineered Quantum Systems, Centre for Quantum Computer and Communication Technology, School of Mathematics and Physics, University of Queensland, Brisbane, Queensland 4072, Australia}

\begin{abstract}
{Measuring entanglement is a demanding task that usually requires full tomography of a quantum system, involving a number of observables that grows exponentially with the number of parties. Recently, it was suggested that adding a single ancillary qubit would allow for the efficient measurement of concurrence, and indeed any entanglement monotone associated to antilinear operations. Here, we report on the experimental implementation of such a device---an embedding quantum simulator---in photonics, encoding the entangling dynamics of a bipartite system into a tripartite one. We show that bipartite concurrence can be efficiently extracted from the measurement of merely two observables, instead of fifteen, without full tomographic information.}
\end{abstract}

\maketitle

Entanglement is arguably the most striking feature of quantum mechanics~\cite{Horodecki1}, defining a threshold between the classical and quantum behavior of nature. Yet its experimental quantification in a given system remains challenging. Several measures of entanglement involve unphysical operations, such as antilinear operations, on the quantum state~\cite{conc:Wootters,Osterloh05}, and thus its direct measurement cannot be implemented in the laboratory. As a consequence, in general, experimental measurements of entanglement have been carried out mostly via the full reconstruction of the quantum state~\cite{mqubits:White}. While this technique---called quantum state tomography~(QST)---has been widely used when dealing with relatively low-dimensional systems~\cite{Haffner05,8qent:Pan}, it is known to become rapidly intractable as the system size grows, being outside of experimental reach in systems with $\sim 10$ qubits~\cite{14qent:Blatt}. This difficulty lies in having to measure an exponentially-growing number of observables, $2^{2N}{-}1$, to reconstruct $N$-qubits. Such constraint can be relaxed somewhat by using, for example, multiple copies of the same quantum state~\cite{entMeas:Walborn}, prior state knowledge in noisy dynamics~\cite{detEnt:Farias}, compressed sensing methods~\cite{sensing:Gross}, or measuring phases monotonically dependent on entanglement~\cite{holphase:Loredo}. However, measuring entanglement in scalable systems remains a challenging task.

An efficient alternative is to embed the system dynamics into an enlarged Hilbert-space simulator, called embedding quantum simulator (EQS)~\cite{DiCandia13, Pedernales14}, where unphysical operations are mapped onto physical transformations on the simulator. The price to pay, comparatively small in larger systems, is the addition of only one ancillary qubit and, usually, dealing with more involved dynamics. However, measuring the entanglement of the simulated system becomes efficient, involving the measurement of a low number of observables in the EQS, in contrast to the $2^{2N}{-}1$ needed with full tomography.

In this Letter, we experimentally demonstrate an embedding quantum simulator, using it to efficiently measure two-qubit entanglement. Our EQS uses three polarization-encoded qubits in a circuit with two concatenated controlled-sign gates. The measurement of only 2 observables on the resulting tripartite state gives rise to the efficient measurement of bipartite concurrence, which would otherwise need 15 observables.
\\

\emph{Protocol.}~We consider the simulation of two-qubit entangling dynamics governed by the Hamiltonian $H{=}{-} g \sigma_z\otimes\sigma_z$, where $\sigma_z{=}\ketbra{0}{0}{-}\ketbra{1}{1}$ is the z-Pauli matrix written in  the computational basis, $\{\ket{0},\ket1\}$, and $g$ is a constant with units of frequency. For simplicity, we let $\hbar{=}1$. Under this Hamiltonian, the concurrence~\cite{conc:Wootters} of an evolving pure state $\ket{\psi(t)}$ is calculated as $\mathcal{C}{=}\left| \bra{\psi(t)}\sigma_y\otimes\sigma_yK\ket{\psi(t)} \right|$, where $K$ is the complex conjugate operator defined as $K\ket{\psi(t)}{=}\ket{\psi(t)^*}$.
\begin{figure}[htp]
\centering
\includegraphics[width=7.5cm]{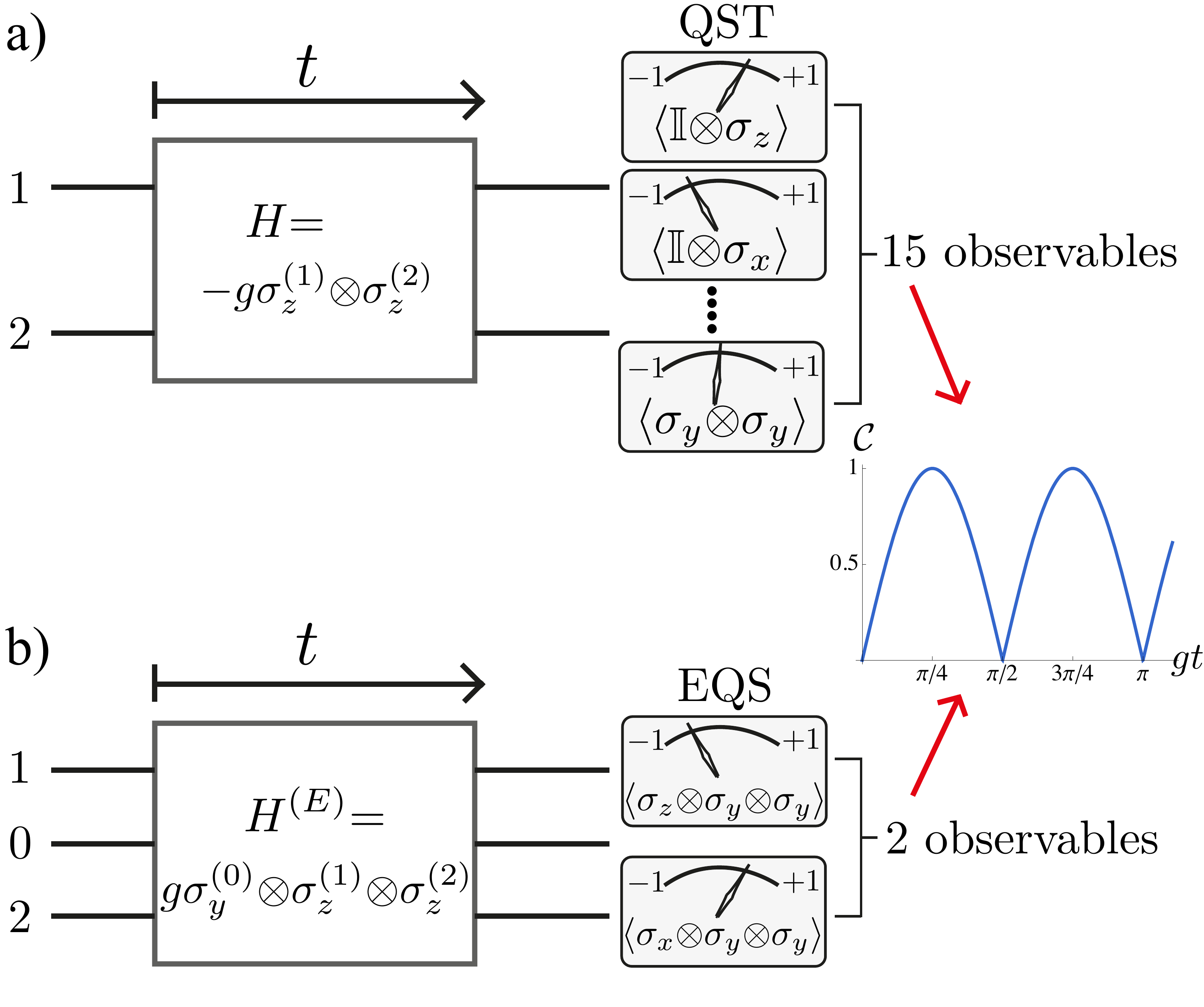}
\caption{(a) Qubits $1$ and $2$ evolve via an entangling Hamiltonian $H$ during a time interval $t$, at which point quantum state tomography (QST) is performed via the measurement of $15$ observables to extract the amount of evolving concurrence. (b) An efficient alternative corresponds to adding one extra ancilla, qubit $0$, and having the enlarged system---the embedding quantum simulator (EQS)---evolve via $H^{(E)}$. Only two observables are now required to reproduce measurements of concurrence of the system under simulation.}
\label{Fig:1}
\end{figure}
Notice here the explicit dependance of $\mathcal C$ upon the unphysical transformation $K$. We now consider the dynamics of the initial state $\ket{\psi(0)}{=}(\ket{0}{+}\ket{1}){\otimes}(\ket{0}{+}\ket{1})/2$. Under these conditions one can calculate the resulting concurrence at any time $t$ as
\begin{equation}
\label{eq:csim}
\mathcal{C}=|\sin(2gt)| .
\end{equation}
The target evolution, $e^{-iHt} \ket{\psi(0)}$, can be embedded in a $3$-qubit simulator. Given the state of interest $\ket{\psi}$, the transformation
\begin{equation}
\label{eq:toSIM}
\ket{\Psi}=\ket{0}\otimes\text{Re}\ket{\psi}+\ket{1}\otimes\text{Im}\ket{\psi},
\end{equation}
gives rise to a real-valued $3$-qubit state $\ket{\Psi}$ in the corresponding embedding quantum simulator. The decoding map is, accordingly, $\ket{\psi}{=}\braket{0}{\Psi}{+}i\braket{1}{\Psi}$. The physical unitary gate $\sigma_z{\otimes}\mathbb{I}_4$ transforms the simulator state into $\sigma_z{\otimes}\mathbb{I}_4\ket\Psi{=}\ket{0}{\otimes}\text{Re}\ket{\psi}{-}\ket{1}{\otimes}\text{Im}\ket{\psi}$, which after the decoding  becomes $\braket{0}{\Psi}{-}i\braket{1}{\Psi}{=}\text{Re}\ket\psi{-}{i}\text{Im}\ket\psi{=}\ket{\psi^*}$. Therefore, the action of the complex conjugate operator $K$ corresponds to the single qubit rotation $\sigma_z{\otimes}\mathbb{I}_4$~\cite{DiCandia13, Casanova11}. Now, following the same encoding rules: $\bra{\psi}OK\ket{\psi}{=}\bra{\Psi}(\sigma_z{-}i\sigma_x){\otimes}O\ket{\Psi}$, with $O$ an observable in the simulation. In the case of  $O{=}\sigma_y{\otimes}\sigma_y$, we obtain
\begin{equation}
\label{eq:C}
\mathcal{C}=|\langle\sigma_z\otimes\sigma_y\otimes\sigma_y\rangle-i\langle\sigma_x\otimes\sigma_y\otimes\sigma_y\rangle|,
\end{equation}
which relates the simulated concurrence to the expectation values of two nonlocal operators in the embedding quantum simulator. Regarding the dynamics, it can be shown that the Hamiltonian $H^{(E)}$ that governs the evolution in the simulator is $H^{(E)}{=}{-}\sigma_y{\otimes}(\text{Re}H){+}i\mathbb{I}_2{\otimes}(\text{Im}H)$~\cite{DiCandia13}. Accordingly, in our case, it will be given by $H^{(E)}{=}g \sigma_y{\otimes}\sigma_z{\otimes}\sigma_z$.

Our initial state under simulation is $\ket{\psi(0)}{=}(\ket{0}{+}\ket{1}){\otimes}(\ket{0}{+}\ket{1})/2$, which requires, see Eq.~(\ref{eq:toSIM}), the initialization of the simulator in $|{\Psi}(0)\rangle{=}|0\rangle{\otimes}\left( |0\rangle+|1\rangle \right){\otimes}\left( |0\rangle+|1\rangle \right){/}2$. Under these conditions, the relevant simulator observables, see Eq.~(\ref{eq:C}), read $\langle\sigma_x{\otimes}\sigma_y{\otimes}\sigma_y\rangle{=}\sin{(2gt)}$ and $\langle\sigma_z{\otimes}\sigma_y{\otimes}\sigma_y\rangle{=}0$, from which the concurrence of Eq.~(\ref{eq:csim}) will be extracted. Therefore, our recipe, depicted in Fig.~\ref{Fig:1}, allows the encoding and efficient measurement of two-qubit concurrence dynamics.

\begin{figure}[h!]
\centering
\includegraphics[width=6.5cm]{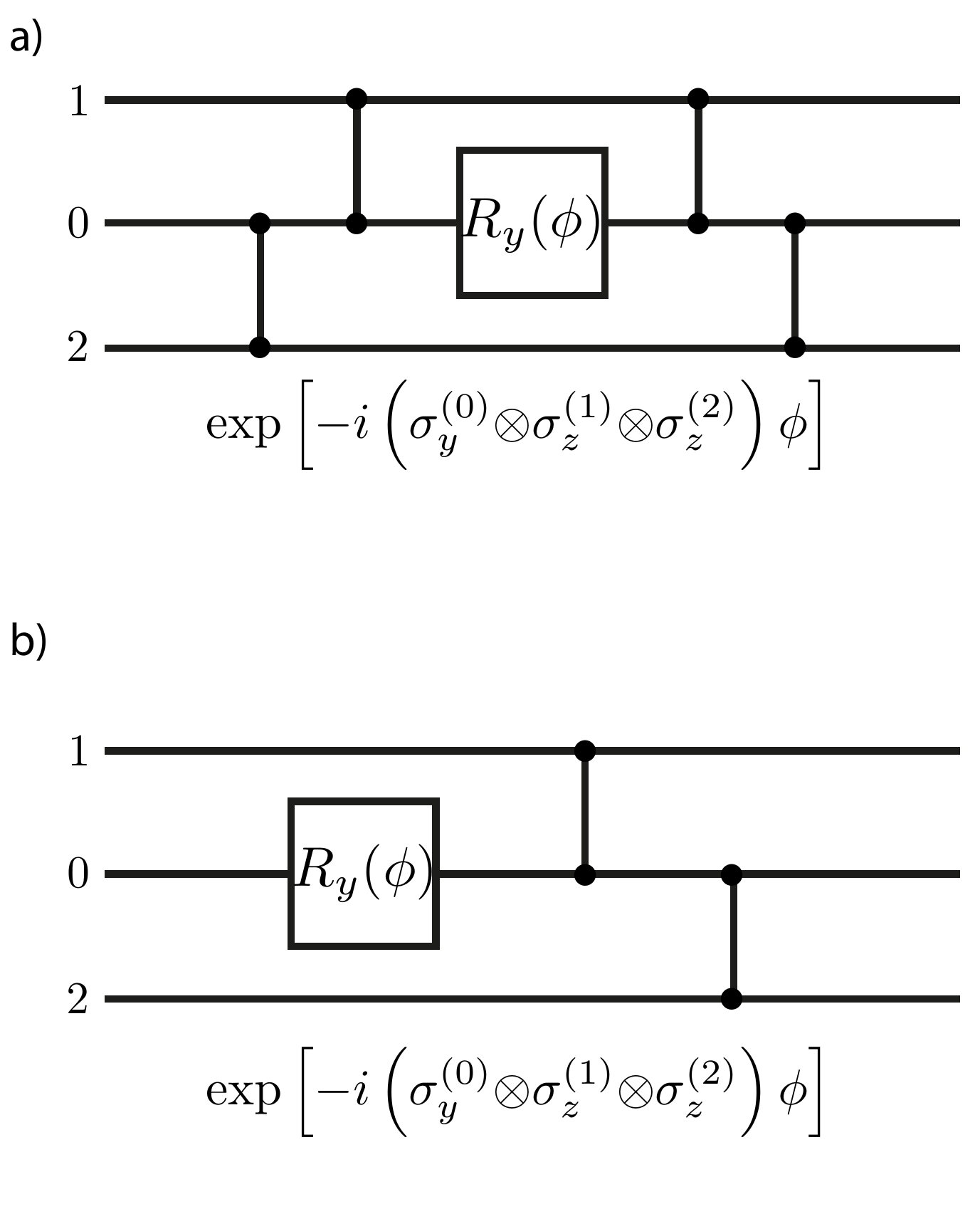}
\caption{Quantum circuit for the embedding quantum simulator. (a) $4$ controlled-sign gates and one local rotation $R_{y}(\phi)$ implement the evolution operator $U({t}){=}\text{exp}\left(-i g \sigma_y^{(0)}{\otimes}\sigma_z^{(1)}{\otimes}\sigma_z^{(2)}t\right)$, with $\phi=gt$. (b) A reduced circuit employing only two controlled-sign gates reproduces the desired three-qubit dynamics for inputs with the ancillary qubit in $|0\rangle$.}
\label{Fig:2}
\end{figure}

To construct the described three-qubit simulator dynamics, it can be shown (see Supplemental Material) that a quantum circuit consisting of $4$ controlled-sign gates and one local rotation $R_y(\phi){=}\text{exp}\left({-i}\sigma_{y} \phi \right)$, as depicted in Fig.~\ref{Fig:2}(a), implements the evolution operator $U(t){=}\text{exp}\left[-ig \left(\sigma_y{\otimes}\sigma_z{\otimes}\sigma_z\right)t\right]$, reproducing the desired dynamics, with $\phi=gt$. This quantum circuit can be further reduced if we consider only inputs with the ancillary qubit in state $| 0 \rangle$, in which case, only two controlled-sign gates reproduce the same evolution, see Fig.~\ref{Fig:2}~(b). This reduced subspace of initial states corresponds to simulated input states of only real components. 

\begin{figure*}[htb]
\centering
\includegraphics[width=13cm]{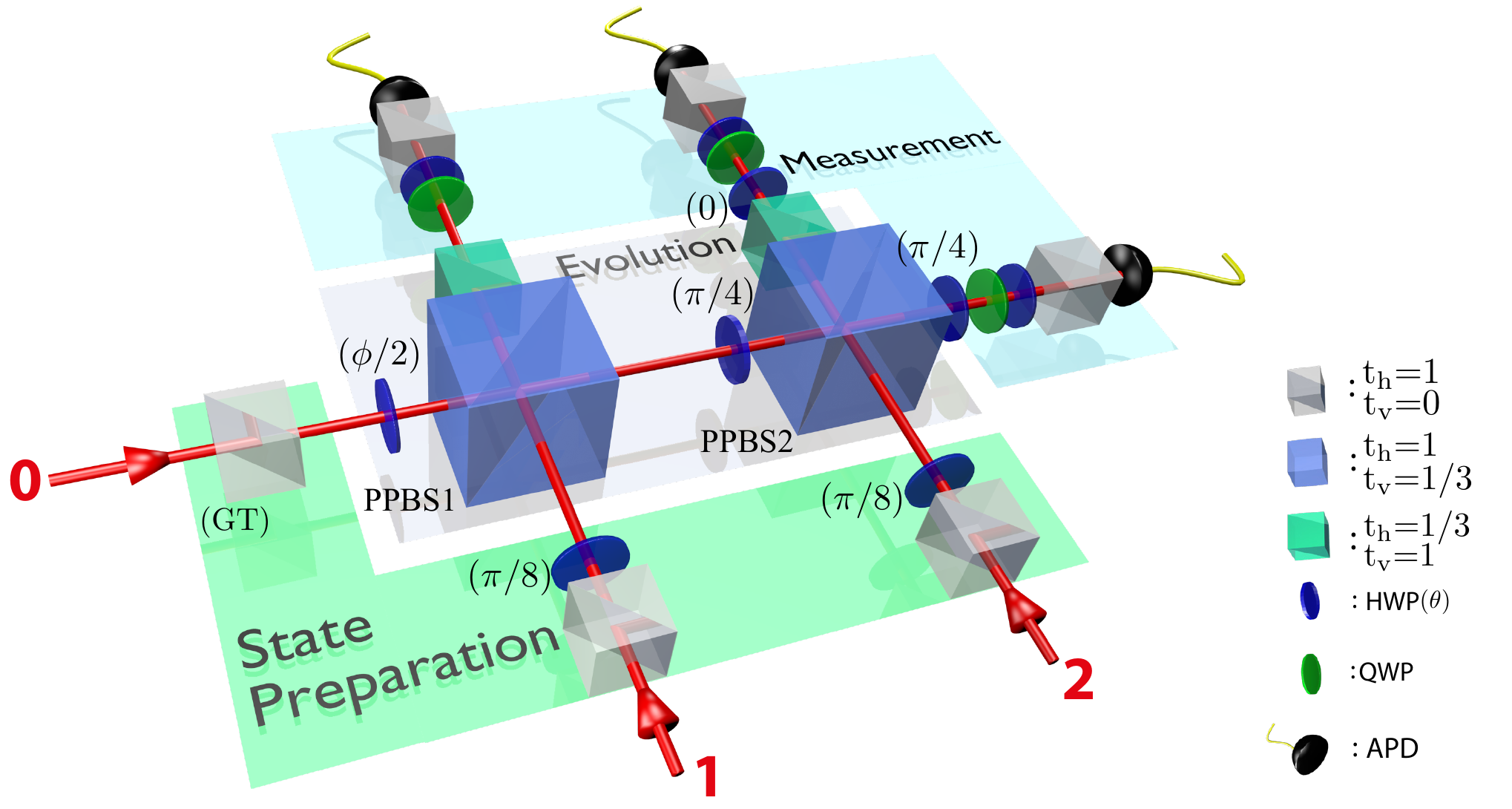}
\caption{Experimental setup. Three single-photons with wavelength centered at 820~nm are injected via single-mode fibers into spatial modes $0$, $1$ and $2$. Glan-Taylor (GT's) prisms, with transmittance $t_\text{h}{=}1$ ($t_\text{v}{=}0$) for horizontal (vertical) polarization, and half-wave plates (HWP's) are employed to initialize the state. Controlled two-qubit operations are performed based on two-photon quantum interference at partially polarizing beam-splitters (PPBS's). Projective  measurements are carried out with a combination of half-wave plates, quarter-wave plates (QWP's) and Glan-Taylor prisms. The photons are collected via single-mode fibers and detected by avalanche photodiodes (APD's).}
\label{Fig:3}
\end{figure*}

\emph{Experimental implementation.}~We encode a three-qubit state in the polarization of 3 single-photons. The logical basis is encoded according to $\ket{h}{\equiv}\ket{0},\ket{v}{\equiv}\ket1$, where $\ket{h}$ and $\ket{v}$ denote horizontal and vertical polarization, respectively. The simulator is initialized in the state $|{\Psi}(0)\rangle{=}|h\rangle^{(0)}{\otimes}\left(|h\rangle^{(1)}+|v\rangle^{(1)}\right){\otimes}\left( |h\rangle^{(2)}+|v\rangle^{(2)}\right){/}2$ of qubits $0$, $1$ and $2$, and evolves via the optical circuit in Fig.~\ref{Fig:2}~(b). Figure~\ref{Fig:3} is the physical implementation of Fig.~\ref{Fig:2}~(b), where the dimensionless parameter $\phi{=}gt$ is controlled by the angle $\phi{/}2$ of one half-wave plate. The two concatenated controlled-sign gates are implemented by probabilistic gates based on two-photon quantum interference~\cite{simpleLO:Ralph,cnot:Obrien,scalecz:Ralph}, see Supplemental Material.

In order
to reconstruct the two three-qubit observables in Eq.~(\ref{eq:C}), one needs to collect $8$ possible tripartite correlations of the observable eigenstates. For instance, the observable $\langle\sigma_x{\otimes}\sigma_y{\otimes}\sigma_y\rangle$ is obtained from measuring the $8$ projection combinations of the $\{\ket{d},\ket{a}\}{\otimes}\{\ket{r},\ket{l}\}{\otimes}\{\ket{r},\ket{l}\}$ states, where $\ket{d}{=}(\ket{h}{+}\ket{v})/\sqrt2$, $\ket{r}{=}(\ket{h}{+}i\ket{v})/\sqrt2$, and $\ket{a}$ and $\ket{l}$ are their orthogonal states, respectively. To implement these polarization projections, we employed Glan-Taylor prisms due to their high extinction ratio. However, only their transmission mode is available, which required each of the $8$ different projection settings separately, extending our data-measuring time. The latter can be avoided by simultaneously registering both outputs of a projective measurement, such as at the two output ports of a polarizing beam splitter, allowing the simultaneous recording of all $8$ possible projection settings. Thus, an immediate reconstruction of each observable is possible.

Our source of single-photons consists of four-photon events collected from the forward and backward pair emission in spontaneous parametric down-conversion in a \emph{beta}-barium borate (BBO) crystal pumped by a $76$~MHz frequency-doubled mode-locked femtosecond Ti:Sapphire laser. One of the four photons is sent directly to an avalanche photodiode detector (APD) to act as a trigger, while the other $3$ photons are used in the protocol. This kind of sources are known to suffer from undesired higher-order photon events that are ultimately responsible of a non-trivial gate performance degradation~\cite{2company:Weinhold,roadTol:Weinhold,multiphEnt:Pan}, although they can be reduced by decreasing the laser pump power. However, given the probabilistic nature and low efficiency of down-conversion processes, multi-photon experiments are importantly limited by low count-rates, see Supplemental Material. Therefore, increasing the simulation performance quality by lowering the pump requires much longer integration times to accumulate meaningful statistics, which ultimately limits the number of measured experimental settings.

As a result of these higher-order noise terms, a simple model can be considered to account for non-perfect input states. The experimental input $n$-qubit state $\rho_{\rm exp}$ can be regarded as consisting of the ideal state $\rho_{\rm id}$ with certain probability $\varepsilon$, and a white-noise contribution with probability $1{-}\varepsilon$, i.e.~$\rho_{\rm exp}{=}\varepsilon \rho_{\rm id}{+}(1{-}\varepsilon)\mathbb{I}_{2^n}{/}{2^n}$.
	\begin{figure*}[htp!]
		\centering
		\includegraphics[width=16cm]{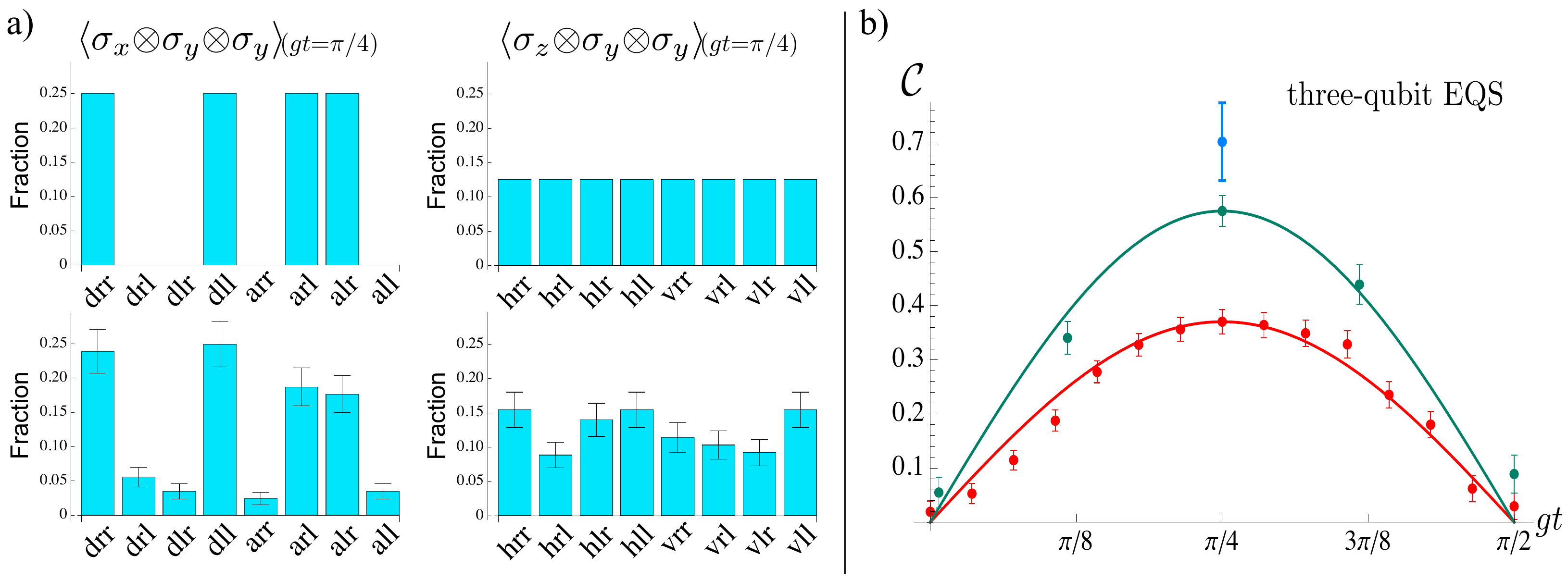} 
		\caption{(a) Theoretical predictions (top) and experimentally measured (bottom) fractions involved in reconstructing $\langle\sigma_x{\otimes}\sigma_y{\otimes}\sigma_y\rangle$ (left) and $\langle\sigma_z{\otimes}\sigma_y{\otimes}\sigma_y\rangle$ (right), taken at $gt{=}\pi{/}4$ for a $10\%$ pump. (b) Extracted simulated concurrence within one evolution cycle, taken at $10\%$ (blue), $30\%$ (green), and $100\%$ (red) pump powers. Curves represent $\mathcal{C}{=}\mathcal{C}_{\text{pp}}|\sin(2gt)|$, where $\mathcal{C}_{\text{pp}}$ is the maximum concurrence extracted for a given pump power (pp): $\mathcal{C}_{10\%}{=}0.70\pm0.07$, $\mathcal{C}_{30\%}{=}0.57\pm0.03$ and $\mathcal{C}_{100\%}{=}0.37\pm0.02$. Errors are estimated from propagated Poissonian statistics. The low count-rates of the protocol, see Supplemental Material, limit the number of measured experimental settings, hence only one data point could be reconstructed at $10\%$ pump.}
		\label{Fig:4}
	\end{figure*}
Since the simulated concurrence is expressed in terms of tensorial products of Pauli matrices, the experimentally simulated concurrence becomes $\mathcal{C}_{\rm exp}{=}\varepsilon|\sin(2gt)|$.

In Fig.~\ref{Fig:4}, we show our main experimental results from our photonic embedding quantum simulator for one cycle of concurrence evolution taken at different pump powers:~$60$~mW, $180$~mW, and $600$~mW---referred as to $10\%$, $30\%$, and $100\%$ pump, respectively. Figure~\ref{Fig:4}~(a) shows theoretical predictions (for ideal pure-state inputs) and measured fractions of the different projections involved in reconstructing $\langle\sigma_x{\otimes}\sigma_y{\otimes}\sigma_y\rangle$ and $\langle\sigma_z{\otimes}\sigma_y{\otimes}\sigma_y\rangle$ for $10\%$ pump at $gt{=}\pi/4$. From measuring these two observables, see Eq.~(\ref{eq:C}), we construct the simulated concurrence produced by our EQS, shown in Fig.~\ref{Fig:4}~(b). We observe a good behavior of the simulated concurrence, which preserves the theoretically predicted sinusoidal form. The overall attenuation of the curve is in agreement with the proposed model of imperfect initial states. Together with the unwanted higher-order terms, we attribute the observed degradation to remaining spectral mismatch between photons created by independent down-conversion events and injected to inputs $0$ and $2$ of Fig.~\ref{Fig:3}---at which outputs $2$~nm band-pass filters with similar but not identical spectra were used.

We compare our measurement of concurrence via our simulator with an explicit measurement from state tomography. In the latter we inject one down-converted pair into modes $0$ and $1$ of Fig.~\ref{Fig:3}. For any value of $t$, set by the wave-plate angle $\phi$, this evolving state has the same amount of concurrence as the one from our simulation, they are equivalent in the sense that one is related to the other at most by local unitaries, which could be seen as incorporated in either the state preparation or within the tomography settings.

\begin{figure}[htb]
		\centering
		\includegraphics[width=7.cm]{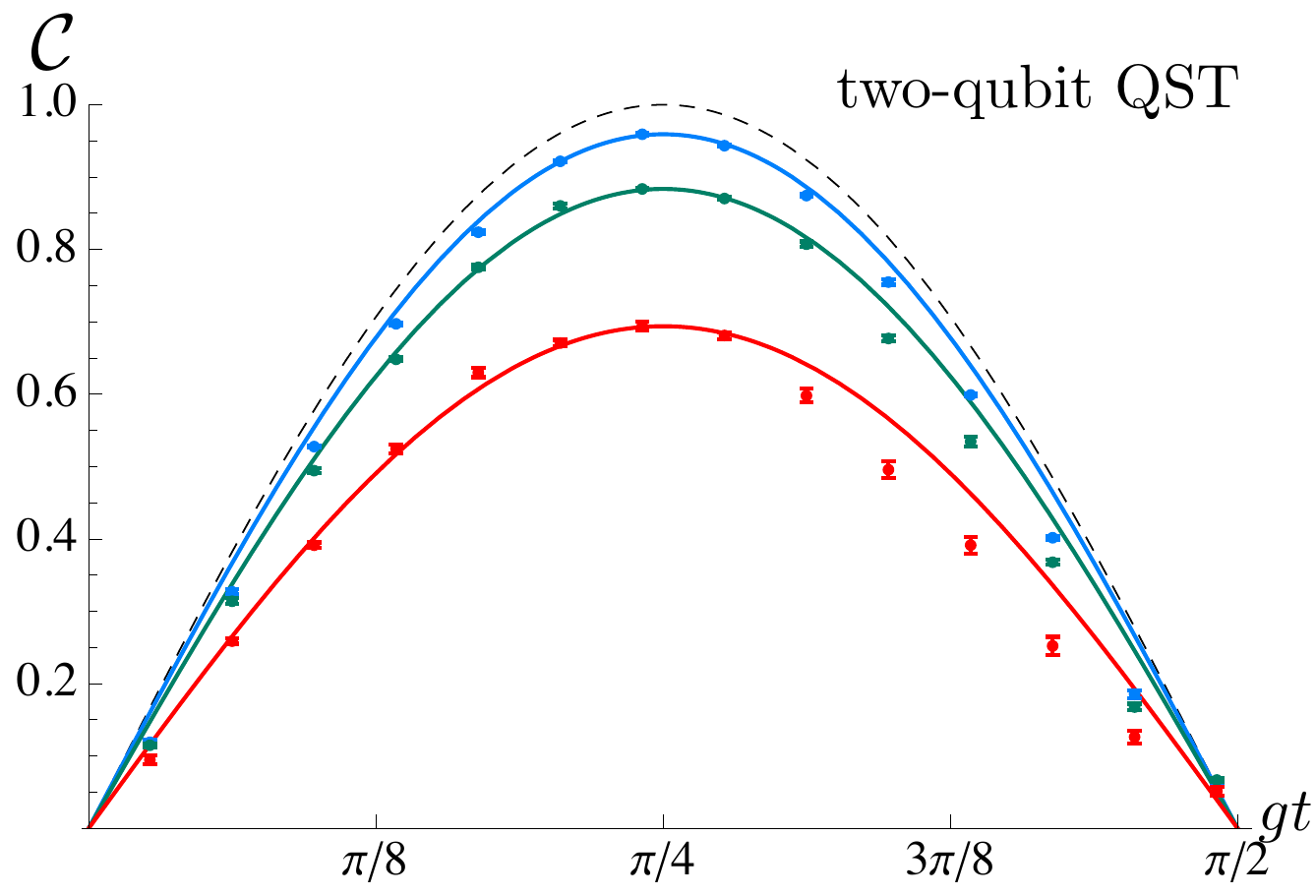}
		\caption{Concurrence measured via two-qubit quantum state tomography (QST) on the explicit two-photon evolution, taken at $10\%$ (blue), $30\%$ (green), and $100\%$ (red) pump powers. The corresponding curves indicate $\mathcal{C}{=}\mathcal{C}_{\text{pp}}|\sin(2gt)|$, with $\mathcal{C}_{\text{pp}}$ the maximum extracted concurrence for a given pump power (pp): $\mathcal{C}_{10\%}{=}0.959\pm0.002$, $\mathcal{C}_{30\%}{=}0.884\pm0.002$, and $\mathcal{C}_{100\%}{=}0.694\pm0.006$. Errors are estimated from Monte-Carlo simulations of Poissonian counting fluctuations.} 
		\label{Fig:5}
	\end{figure}

Figure~\ref{Fig:5} shows our experimental results for the described two-photon protocol. We extracted the concurrence of the evolving two-qubit state from overcomplete measurements in quantum state tomography~\cite{mqubits:White}.
A maximum concurrence value of $\mathcal{C}{=}1$ is predicted in the ideal case of perfect pure-state inputs. Experimentally, we measured maximum values of concurrence of $\mathcal{C}_{10\%}{=}0.959\pm0.002$, $\mathcal{C}_{30\%}{=}0.884\pm0.002$ and $\mathcal{C}_{100\%}{=}0.694\pm0.006$, for the three different pump powers, respectively. For the purpose of comparing this two-photon protocol with our embedding quantum simulator, only results for the above mentioned powers are shown. However, we performed an additional two-photon protocol run at an even lower pump power of $30$~mW ($5\%$ pump), and extracted a maximum concurrence of $\mathcal{C}_{5\%}{=}0.979\pm0.001$. A clear and pronounced decline on the extracted concurrence at higher powers is also observed in this protocol. However, a condition closer to the ideal one is reached. This observed pump power behavior and the high amount of measured concurrence suggest a high-quality gate performance, and that higher-order terms---larger for higher pump powers---are indeed the main cause of performance degradation.

While only mixed states are always involved in experiments, different degrees of mixtures are present in the $3$- and $2$-qubit protocols, resulting in different extracted concurrence from both methods. An inspection of the pump-dependence, see Supplemental Material, reveals that both methods decrease similarly with pump power and are close to performance saturation at the $10\%$ pump level. This indicates that in the limit of low higher-order emission our $3$-qubit simulator is bounded to the observed performance. Temporal overlap between the $3$ photons was carefully matched. Therefore, we attribute the remaining discrepancy to spectral mismatch between photons originated from independent down-conversion events. This disagreement can in principle be reduced via error correction~\cite{errC:Brien,errC:Pan} and entanglement purification~\cite{entPur:Pan} schemes with linear optics.

\emph{Discussion.}~We have shown experimentally that entanglement measurements in a quantum system can be efficiently done in a higher-dimensional embedding quantum simulator. The manipulation of larger Hilbert spaces for simplifying the processing of quantum information has been previously considered~\cite{simpLogic:Lanyon}. However, in the present scenario, this advantage in computing concurrence originates from higher-order quantum correlations, as it is the case of the appearance of tripartite entanglement~\cite{shor:Pan,shor:White}.

The efficient behavior of embedding quantum simulators resides in reducing an exponentially-growing number of observables to only a handful of them for the extraction of entanglement monotones. We note that in this non-scalable photonic platform the addition of one ancillary qubit and one entangling gate results in count rates orders of magnitude lower as compared to direct state tomography on the $2$-qubit dynamics. This means that in practice absolute integration times favor the direct $2$-qubit implementation. However, this introduced limitation escapes from the purposes of the embedding protocol and instead belongs to the specific technology employed in its current state-of-the-art performance.

This work represents the first proof-of-principle experiment showing the efficient behavior of embedding quantum simulators for the processing of quantum information and extraction of entanglement monotones. This validates an architecture-independent paradigm that, when implemented in a scalable platform, e.g. ions~\cite{14qent:Blatt,Pedernales14}, would overcome a major obstacle in the characterization of large quantum systems. The relevance of these techniques will thus become patent as quantum simulators grow in size and currently standard approaches like full tomography become utterly unfeasible. We believe that these results pave the way to the efficient measurement of entanglement in any quantum platform via embedding quantum simulators.

We thank M.~A.~Broome for helpful discussions. This work was supported by the Centre for Engineered Quantum Systems (Grant No. CE110001013) and the Centre for Quantum Computation and Communication Technology (Grant No. CE110001027). M.~P.~A. acknowledges support from the Australian Research Council Discovery Early Career Awards (No. DE120101899). A.~G.~W. was supported by the University of Queensland Vice-Chancellor's Senior Research Fellowship. J. C. acknowledge support from the Alexander von Humboldt Foundation, while R.~D.~C., J.~S.~P., and E.~S. from Basque Government IT472-10; Spanish MINECO FIS2012-36673-C03-02; UPV/EHU UFI 11/55; UPV/EHU PhD fellowship; PROMISCE and SCALEQIT EU projects.
\\
\\
\emph{Note added.}---We recently learned of a related paper by Chen et al~\cite{Chen:EQS}.

\section{Supplemental Material}
\subsection{I.~Quantum circuit of the embedding quantum simulator}

Following the main text, the evolution operator associated with the embedding Hamiltonian $H^{(E)}{=}g\sigma_y{\otimes}\sigma_z {\otimes}\sigma_z$ can be implemented via 4 control-$Z$ gates ($CZ$), and a single qubit rotation $R_y(t)$. These gates act as
\begin{eqnarray}
CZ^{ij}&=&|0\rangle\langle0|^{(i)}\otimes \mathbb{I}^{(j)}+|1\rangle\langle1|^{(i)}\otimes \sigma^{(j)}_z,\\
R^i_y(t)&=&e^{-i\sigma_y^{(i)}gt}\equiv (\cos(gt)\mathbb{I}^{(i)}-i\sin(gt)\sigma_y^{(i)}),
\end{eqnarray}
with $\sigma_z{=}|0\rangle\langle0|{-}|1\rangle\langle1|$, and $\sigma_y{=}-i|0\rangle\langle1|{+}i|1\rangle\langle0|$. The indices $i$ and $j$ indicate on which particle the operators act. The circuit for the embedding quantum simulator consists of a sequence of gates applied in the following order:
	\begin{equation}
		U(t)=CZ^{02}CZ^{01}R^0_y(t)CZ^{01}CZ^{02}.
	\end{equation}
Simple algebra shows that this expression can be recast as
\begin{eqnarray}\nonumber
U(t)&=& \cos (gt) \mathbb{I}^{(0)} \otimes \mathbb{I}^{(1)} \otimes \mathbb{I}^{(2)} - i \sin (gt) \sigma_y^{(0)} \otimes \sigma_z^{(1)}\otimes \sigma_z^{(2)}\\ 
&=& \text{exp}\left({-i g\sigma_y^{(0)} \otimes \sigma_z^{(1)}\otimes \sigma_z^{(2)} t}\right),
\end{eqnarray} 
explicitly exhibiting the equivalence between the gate sequence and the evolution under the Hamiltonian of interest. 

\subsection{II. Linear optics implementation}
\begin{figure}[htp]
		\begin{center}
		\includegraphics [width= 8.cm]{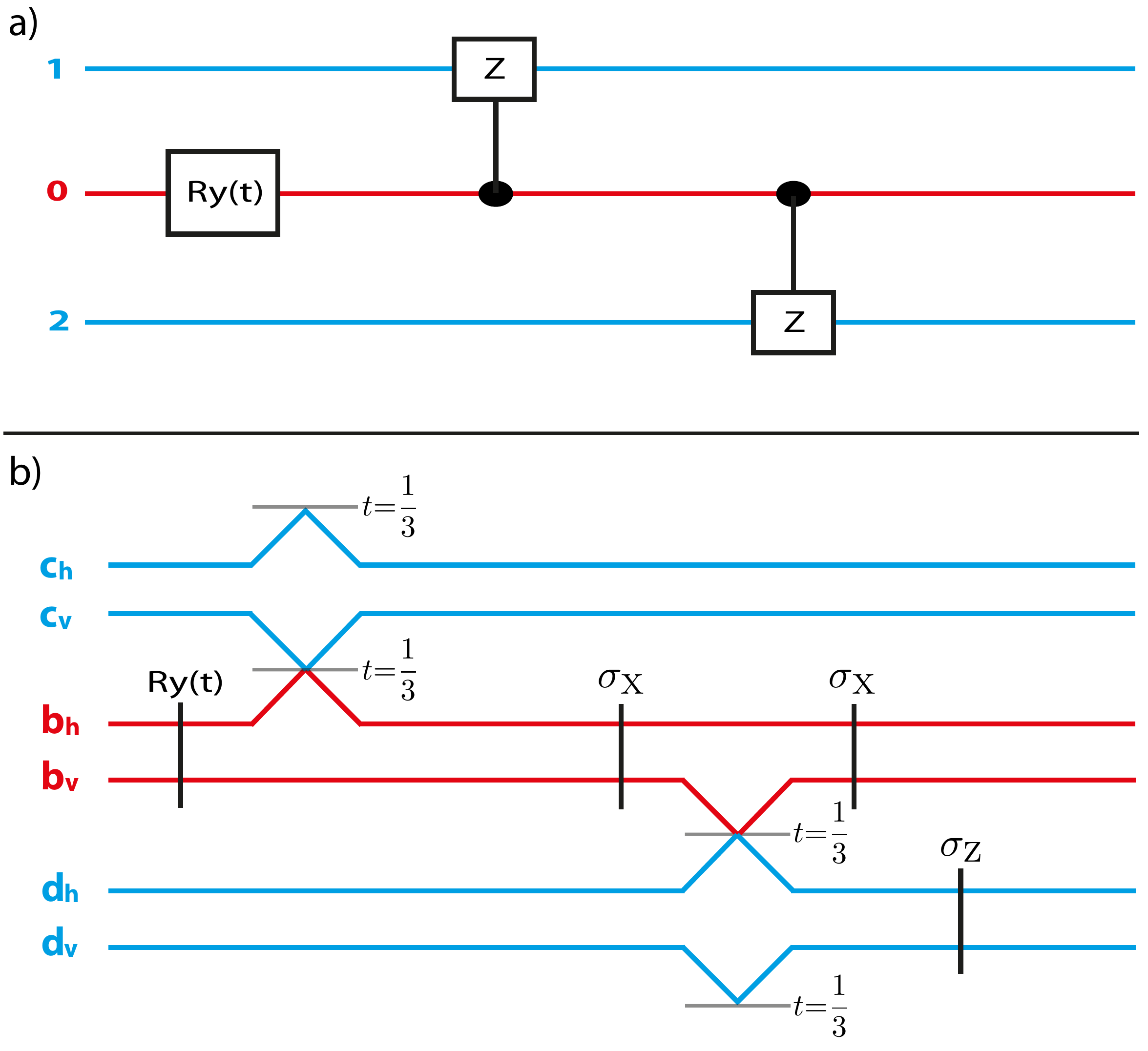}
		\end{center}
\caption{(a)~Circuit implementing the evolution operator $U(t){=}\text{exp}\left({-i g\sigma_y^{(0)}{\otimes}\sigma_z^{(1)}{\otimes}\sigma_z^{(2)} t}\right)$, if the initial state is $|{\Psi}(0)\rangle{=}|0\rangle^{(0)}{\otimes}\left( |0\rangle^{(1)}+|1\rangle^{(1)} \right){\otimes}\left( |0\rangle^{(2)}+|1\rangle^{(2)} \right){/}2$. (b)~Dual-rail representation of the circuit implemented with linear-optics. Red (blue) lines represent trajectories undertaken by the control qubit (target qubits).}\label{fig:SM1}
	\end{figure}

The evolution of the reduced circuit is given by a $R_y(t)$ rotation of qubit $0$, followed by two consecutive control-Z gates on qubits $1$ and $2$, both controlled on qubit $0$, see Fig.~\ref{fig:SM1}~(a). These logic operations are experimentally implemented by devices that change the polarization of the photons, where the qubits are encoded, with transformations as depicted in Fig.~\ref{fig:SM1}~(b). For single qubit rotations, we make use of half-wave plates (HWP's), which shift the linear polarization of photons. For the two-qubit gates, we make use of two kinds of partially-polarizing beam splitters (PPBS's). PPBS's of type $1$ have transmittances $t_h{=}1$ and $t_v{=}1{/}3$ for horizontal and vertical polarizations, respectively. PPBS's of type $2$, on the other hand, have transmittances $t_h{=}1{/}3$ and $t_v{=}1$. Their effect can be expressed in terms of polarization dependant input-output relations---with the transmitted mode corresponding to the output mode---of the bosonic creation operators as
\begin{align}\label{eq:bsaout}
a_{p,out}^{\dag(i)}&=\sqrt{t_{p}}a_{p,in}^{\dag(i)}+\sqrt{1-t_p}a_{p,in}^{\dag(j)}\\ \label{eq:bsbout}
a_{p,out}^{\dag(j)}&=\sqrt{1-t_p}a_{p,in}^{\dag(i)}-\sqrt{t_p}a_{p,in}^{\dag(j)},
\end{align}
where $a_{p,in}^{\dag(i)}$ ($a_{p,out}^{\dag(i)}$) stands for the $i$-th input (output) port of a PPBS with transmittance $t_p$ for $p$-polarized photons. 

Our circuit is implemented as follows: the first $R_y(t)$ rotation is implemented via a HWP oriented at an angle $\theta=gt{/}2$ with respect to its optical axis. The rest of the target circuit, corresponding to the sequence of two control-Z gates, can be expressed in terms of the transformation of the input to output creation operators as
	\begin{eqnarray}\label{tr:C1}\nonumber
{b_h} {c_h} {d_h} &\to& {b_h} {c_h} {d_h} \\ \nonumber
{b_h} {c_h} {d_v} &\to&{b_h} {c_h} {d_v} \\  \nonumber
{b_h} {c_v} {d_h}&\to&{b_h} {c_v} {d_h}\\  \nonumber
{b_h} {c_v} {d_v}&\to&{b_h} {c_v} {d_v}\\  \nonumber
{b_v} {c_h} {d_h}&\to&{b_v} {c_h} {d_h}\\  \nonumber
{b_v} {c_h} {d_v}&\to&-{b_v} {c_h} {d_v}\\  \nonumber
{b_v} {c_v} {d_h}&\to&-{b_v} {c_v} {d_h}\\ 
{b_v} {c_v} {d_v}&\to&{b_v} {c_v} {d_v},
  	\end{eqnarray}
where $b{\equiv}a^{\dag(0)}$, $c{\equiv}a^{\dag(1)}$, and $d{\equiv}a^{\dag(2)}$ denote the creation operators acting on qubits $0$, $1$, and $2$, respectively. These polarization transformations can be implemented with a probability of $1{/}27$ via a $3$-fold coincidence detection in the circuit depicted in Fig.~\ref{fig:SM1}~(b). In this dual-rail representation of the circuit, interactions of modes $c$ and $d$ with vacuum modes are left implicit.

The $\sigma_x$ and $\sigma_z$ single qubit gates in Fig.~\ref{fig:SM1}~(b) are implemented by HWP's with angles $\pi{/}4$ and $0$, respectively. In terms of bosonic operators, these gates imply the following transformations,
\begin{eqnarray}
\sigma_x:&& \ {b_h}\to{b_v},\quad  \ {b_v}\to{b_h}\\
\sigma_z:&& \  \ {d_h}\to{d_h},\quad {d_v}\to-{d_v}.
\end{eqnarray}

According to all the input-output relations involved, it can be calculated that the optical elements in Fig.~\ref{fig:SM1}~(b) implement the following transformations
	\begin{eqnarray}\label{tr:C1}\nonumber
{b_h} {c_h} {d_h} &\to& {b_h} {c_h} {d_h}/(3\sqrt3) \\ \nonumber
{b_h} {c_h} {d_v} &\to&{b_h} {c_h} {d_v}/(3\sqrt3) \\  \nonumber
{b_h} {c_v} {d_h}&\to&{b_h} {c_v} {d_h}/(3\sqrt3)\\  \nonumber
{b_h} {c_v} {d_v}&\to&{b_h} {c_v} {d_v}/(3\sqrt3)\\  \nonumber
{b_v} {c_h} {d_h}&\to&{b_v} {c_h} {d_h}/(3\sqrt3)\\  \nonumber
{b_v} {c_h} {d_v}&\to&-{b_v} {c_h} {d_v}/(3\sqrt3)\\  \nonumber
{b_v} {c_v} {d_h}&\to&-{b_v} {c_v} {d_h}/(3\sqrt3)\\ 
{b_v} {c_v} {d_v}&\to&{b_v} {c_v} {d_v}/(3\sqrt3),
  	\end{eqnarray}
if events with $0$ photons in some of the three output lines of the circuit are discarded. Thus, this linear optics implementation corresponds to the evolution of interest with success probability $P=(1/(3\sqrt3))^2=1/27$.

\subsection{III. Photon count-rates}

Given the probabilistic nature and low efficiency of down-conversion processes, multi-photon experiments are importantly limited by low count-rates. In our case, typical two-photon rates from source are around ${150}$~kHz at $100\%$ pump (two-photon rates are approx. linear with pump power), which after setup transmission (${\sim}80\%$) and $1{/}9$ success probability of one controlled-sign gate, are reduced to about $13$~kHz ($1$~kHz) at $100\%$ ($10\%$) pump. These count-rates make it possible to run the two-photon protocol, described in the main text, at low powers in a reasonable amount of time. However, this situation is drastically different in the three-photon protocol, where we start with $500$~Hz of $4$-fold events from the source, in which case after setup transmission, $1{/}27$ success probability of two gates, and $50\%$ transmission in each of two $2$~nm filters used for this case, we are left with as few as ${\sim}100$~mHz (${\sim}1$~mHz) at $100\%$ ($10\%$) pump ($4$-fold events reduce quadratically with pump). Consequently, long integration times are needed to accumulate meaningful statistics, imposing a limit in the number of measured experimental settings.
\subsection{IV. Pump-dependence}

	\begin{figure}[htp]
		\begin{center}
		\includegraphics [width= 8.cm]{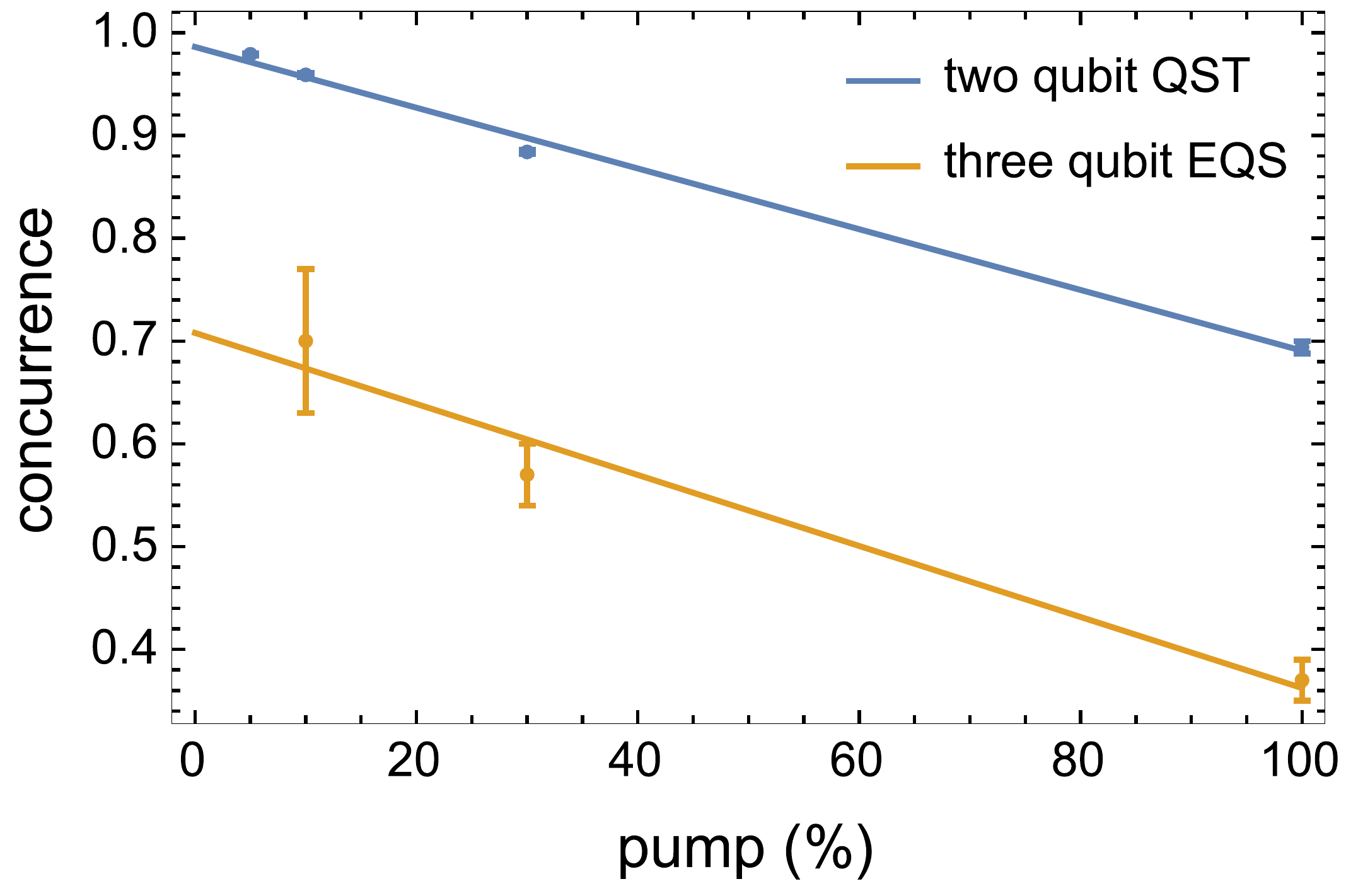}
		\end{center}
\caption{Measured concurrence vs pump power. The concurrence is extracted from both two-qubit quantum state tomography (QST) and the three-qubit embedding quantum simulator (EQS). Straight lines are linear fits to the data. Slopes overlapping within error, namely $-0.0030\pm0.0001$ from QST and $-0.0035\pm0.0007$ from EQS, show that both methods are affected by higher-order terms at the same rate.}\label{fig:SM2}
	\end{figure}
To estimate the effect of power-dependent higher-order terms in the performance of our protocols, we inspect the pump power dependence of extracted concurrence from both methods. Fig.~\ref{fig:SM2} shows that the performances of both protocols decrease at roughly the same rate with increasing pump power, indicating that in both methods the extracted concurrence at $10\%$ pump is close to performance saturation.

\begin{figure}[htp]
		\begin{center}
		\includegraphics [width= 8.cm]{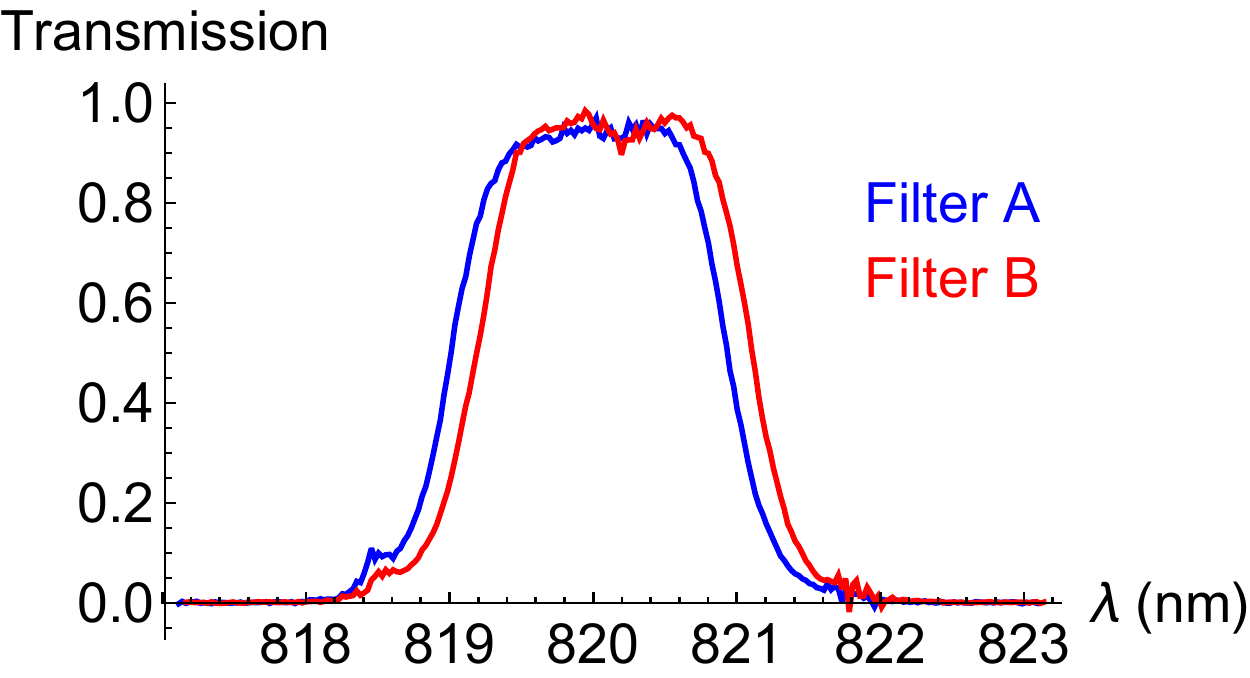}
		\end{center}
\caption{Spectral filtering. The measured transmission for both filters used in our three-qubit protocol qualitatively reveals the remaining spectral mismatch.}\label{fig:SM3}
	\end{figure}

The principal difference between the two methods is that in the three-qubit protocol one of the photons originates from an independent down-conversion event and as such will present a slightly different spectral shape due to a difficulty in optimizing the phase-matching condition for both forward and backward directions simultaneously. To reduce this spectral mismatch, we used two $2$~nm filters at the output of the two spatial modes where interference from independent events occurs, see Fig.~\ref{fig:SM3}. Note that not identical spectra are observed. This limitation would be avoided with a source that presented simultaneous high indistinguishability between all interfering photons.

\end{document}